\documentclass[12pt,preprint]{aastex}

\shorttitle{UV/EUV High-Throughput Spectroscopic Telescope}
\shortauthors{Imada et al.}

\begin{document}

\title{UV/EUV High-Throughput Spectroscopic Telescope: A Next Generation Solar Physics Mission white paper}

\author{S. \textsc{Imada},\altaffilmark{1} 
T. \textsc{Shimizu} \altaffilmark{2}
T. \textsc{Kawate} \altaffilmark{2}
H.  \textsc{Hara},\altaffilmark{3} 
T.  \textsc{Watanabe},\altaffilmark{3}
}
  %%\textsc{Asai},\altaffilmark{2} T.
  %%\textsc{Minoshima},\altaffilmark{3} L. K.
  %%\textsc{Harra},\altaffilmark{4} J. T.
  %%\texts{Mariska}\altaffilmark{5}}
  
\altaffiltext{1}{ Institute for Space-Earth Environmental Research (ISEE), Nagoya University, Furo-cho, Chikusa-ku, Nagoya 464-8601, Japan}
\altaffiltext{2}{Institute of Space and Astronautical Science, Japan Aerospace Exploration Agency, 3--1--1 Yoshinodai, Sagamihara-shi, Kanagawa 229--8510, Japan}
\altaffiltext{3}{ National Astronomical Observatory of Japan,  2--21--1 Osawa, Mitaka-shi, Tokyo 181--8588, Japan}

%\email{shinsuke.imada@nao.ac.jp}
%%\altaffiltext{2}{Nobeyama Solar Radio Observatory, Minamisaku,
%%  Nagano 384--1305, Japan} \altaffiltext{3}{Department of Earth
%%  and Planetary Science, University of Tokyo, 7--3--1 Hongo,
%%  Bunkyo-ku, Tokyo 113--0033, Japan }
%%\altaffiltext{4}{UCL-Mullard Space Science Laboratory, Holmbury
%%  St Mary, Dorking, Surrey, RH5 6NT, UK} \altaffiltext{5}{Space
%%  Science Division, Naval Research Laboratory, Washington DC
%%  20375, USA}

\begin{abstract}
Main Science Target: Quantitatively estimate the energy transfer from the photosphere to the corona, and reveal the plasma dynamics in the vicinity of the energy dissipation region, where is the key region for coronal heating, solar wind acceleration, and/or solar flare, by the high spatial and temporal resolution UV/EUV spectroscopy.
\end{abstract}


\section{Science Backgrounds}

The origin of the activity in the solar corona is a long-standing problem in solar physics. So far, many studies have been devoted toward understanding the origin of the activity in the solar atmosphere. Nowadays, it is widely accepted that the magnetic fields have a key role for the activities in the solar atmosphere. Recent satellite observations, such as Hinode, Solar Dynamics Observatory (SDO), Interface Region Imaging Spectrograph (IRIS), show the detail characteristics of the solar atmosphere and try to reveal the energy transfer from the photosphere to the corona through the magnetic fields and its energy conversion by various processes. However, quantitative estimation of energy transfer along the magnetic field is not enough. There are mainly two reason why it is difficult to observe the energy transfer from photosphere to corona; 1) spatial resolution gap between photosphere (a few 0.1 arcsec) and corona (a few arcsec), 2) lack in temperature coverage. Furthermore, there is not enough observational knowledge of the physical parameters in the energy dissipation region. There are mainly three reason why it is difficult to observe in the vicinity of the energy dissipation region; 1) small spatial scale, 2) short time scale, 3) low emission. It is generally believed that the energy dissipation occurs in the very small scale and its duration is very short ($\sim$10 second). Further, the density in the dissipation region might be very low. Therefore, the high spatial and temporal resolution UV/EUV spectroscopic observation with wide temperature coverage is crucial to estimate the energy transport from photosphere to corona quantitatively and diagnose the plasma dynamics in the vicinity of the energy dissipation region. Below is the example of science objectives. Because of the page limitation, we concentrate on the discussion about diagnostics near the energy dissipation region.
 
\section{The Science Objects}

\subsection{Coronal Heating}

Coronal heating is one of the main problems in the solar physics.
The plausible mechanisms producing the hot tenuous plasma making up the solar corona are still not clear.
It is believed that energy release impulsively occurs in the corona by magnetic reconnection and/or by wave dissipation.
Very hot and tenuous plasma is expected to be generated inside the energy dissipation region.
Then, the thermal conduction quickly conveys the heats from corona to chromosphere and evaporates chromospheric plasma into corona.
The very hot tenuous plasma might be observed only during very short period ($\sim$10 sec).
To diagnose it we need to scan a region about 10000 km wide in a time significantly shorter than the sound speed in ~10 second to follow the conduction evaporation@flow as it moves through the corona.
The very hot and tenuous component in the solar corona has been inferred from X-ray observation (Reale et al. 2009). On the other hand, recent spectroscopic observation can not confirmed the sufficient presence of very hot component \citep{win}.
\cite{ish} tried to confirm the very hot tenuous plasma component by the combination of Hinode EIS, XRT, and FOXSI observation, and found the contribution from 10 MK is very small in differential emission measure.
Actually, these analyses are based on the assumption of the ionization equilibrium.
As mentioned above, these phenomena might occur within very short time and low-density condition.
Thus it is plausible that the ionization cannot reach to the equilibrium stage \citep{bra}.
The interpretation of the very hot component observation might be changed when we consider the equilibrium ionization \citep{ima,ima2}.
To diagnose the plasma dynamics in the vicinity of the energy dissipation region, we have to consider non-equilibrium ionization plasma.
Multi-lines analysis have strong advantage to understand the ionization conditions, because generally different atomic species, for example O, Ca, Si and Fe, have different equilibrium timescale for ionization \citep{rea2}.
By using this information, we can obtain a time history of heating in corona.
This analysis is very important to identify the region where strong heating occur (for example, loop top heating? Or foot point heating?).

Also, very recent high spatial resolution observations (Hi-C; $\sim$0.25 arcsec) show the fine structures which might be associated with current sheet. We can clearly see that most of the coronal structures appear to be resolved at Hi-C spatial resolution. Because as mentioned above the dissipation region is very limited, their observable signals become weaker if we use lower resolution observation (contamination of ambient plasma signals). Therefore, we can conclude that the high spatial ($\sim$0.3 arcsec) and temporal ($\sim$10 second) resolution UV/EUV spectroscopic observation with wide wavelength (temperature) coverage is crucial to understand the key region of coronal heating.

\subsection{Solar Flare}
The recent observations reveal many aspects of solar flares and confirm that the magnetic
reconnection is one of the main energy conversion mechanisms during flares \citep{tsu,mck,sav}. So far, plasma dynamics before and after flare can be discussed in detail from the observations. On the other hand, there is not enough observational knowledge of the physical parameters in the reconnection region. The inflow into the reconnection region, the heating by shocks, and the temperatures and densities of the outflow jets predicted by reconnection, have not been quantitatively measured in sufficient \citep{har,tak,ima3}. One of the reasons why it is difficult to observe inside the reconnection region is due to its faintness. Because the post flare loops are very bright, the magnetic reconnection region is much more hard to observe after the impulsive phase. Thus, we need to observe the reconnection region before growing the bright post flare loops. High-throughput (high temporal resolution) UV/EUV spectrometer is essential to observe the reconnection region. Further, the magnetic reconnection that causes rapid heating by slow-mode shocks might result in a transient ionization condition. High-throughput UV/EUV and wide wavelength coverage spectrometer enables us to observe the transient state of ionization, which tells us a time history of heating by shocks \citep{ima4}. Probing the magnetic reconnection region from the viewpoint of non-equilibrium ionization is very important. Therefore, we need the high temporal resolution ($\sim$10 second) with wide wavelength coverage (multi-lines) UV/EUV spectroscopic observation to understand the key region for solar flare.
Theoretical model of steady fast reconnection, called Petchek-type reconnection, expected two slow-mode shocks elongated from the reconnection region, which play a role in heating ambient plasma \citep{pet}. Recently, the other theoretical model, plasmoid-unstable reconnection, is also discussed extensively \citep{lou}. However, we still do not have a clear answer to which reconnection model is likely in high Lundquist number (collisionless) regime. This is the fundamental and essential problem in not only solar physics but also space plasma physics. The solar flare observation has a great advantage to answer this question, and it is key to diagnose plasma dynamics in the vicinity of magnetic reconnection region.

\section{Key Requirement from the Science Objectives}

The high spatial and temporal resolution UV/EUV spectroscopic observation from space is crucial to estimate the energy transport from photosphere to corona quantitatively and diagnose the plasma dynamics in the vicinity of the energy dissipation region. Because it is necessary to temporally resolve the typical coronal loop length (a few 10000 km) less than the Alfven wave and/or the sound wave (a few 1000 km/sec) traveling time, we need 10 second temporal resolution for scanning the typical scale length. For spatial resolution we need to resolve the fundamental structure, which seems to be ~0.3 arcsec, in the corona. Then the requirement for the exposure time can be estimated ~0.3 second (10 sec / 10 arcsec * 0.3 arcsec ). Below is the summary for what we require to observe the Science Objectives.

\begin{table*}
\begin{center}
\caption{Key Requirements. }
\begin{tabular}{cr}
\tableline\tableline
 & Required Value  \\
\tableline
Spatial resolution & $\sim$0.3 arcsec\\
Typical exposure times & 0.3 second \\
Temperature coverage & 0.01 to 20 MK  \\
Spectral resolution & $\sim$ 20000 \\
Field of View & 300 $\times$ 300 arcsec$^2$ \\
\tableline
\end{tabular}
%% Any table notes must follow the \end{tabular} command.
%\tablecomments{}
\end{center}
\end{table*}

\end{document}